\documentclass[conference]{IEEEtran}

\usepackage{amsmath,amsfonts}

\usepackage{subcaption}
\newsavebox{\mybox}

\usepackage{amsthm}
\usepackage{mdframed}
\usepackage{centernot}
\usepackage{comment}
\usepackage{hyperref}
\usepackage{graphicx}
\usepackage{booktabs}

\newcommand{\Hh}{\mathcal{H}}

\newcommand{\Pp}{\mathcal{P}}

\newcommand{\pP}{\mathbb{P}}

\newcommand{\Dd}{\mathcal{D}}

\newcommand\vd[2]{d_{i, p}}
\newcommand{\set}[1]{\left\{ #1 \right\}}

\newcommand{\R}{\mathbb R}

\newcommand{\Real}{\R}

\newcommand{\toolname}{\textsc{TenForty}\xspace}
\usepackage{tikz}

\definecolor{gold}{rgb}{0.99,0.78,0.07}
\usetikzlibrary{arrows,shapes,snakes,automata,backgrounds,positioning,decorations.pathmorphing,
	decorations.markings,calc}

\tikzstyle{dtreenode}=[draw=blue!10!gray,rounded rectangle, minimum size=5mm,fill=blue!10!white]
\tikzstyle{dtreeleaf}=[draw=black!60,minimum width=1cm,minimum height=0.4cm,rectangle,fill=blue!50!white]
\tikzset{every loop/.style={looseness=7}}
\tikzset{
	gluon/.style={decorate,draw=black,
		decoration={coil,amplitude=1pt, segment length=5pt}}
}
\tikzset{
	gluon1/.style={decorate,draw=black,
		decoration={coil,amplitude=3pt, segment length=3pt}}
}
\tikzset{
	gluonew/.style={decorate,draw=black,
		decoration={coil,amplitude=1pt, segment length=2pt}}
}

\tikzset{bicolor/.style args={#1 and #2 and #3}{
		path picture={
			\tikzset{rounded corners=0}
			\fill [#1] (path picture bounding box.south west)
			rectangle
			($(path picture  bounding box.north west)!#3!(path picture bounding
			box.north east)$);
			\fill [#2]
			($(path picture bounding box.south west)!#3!(path picture bounding
			box.south east)$)
			rectangle (path picture bounding box.north east);
}}}

\tikzset{tricolor/.style args={#1 and #2 and #3 and #4 and #5}{
		path picture={
			\tikzset{rounded corners=0}
			\fill [#1] (path picture bounding box.south west)
			rectangle
			($(path picture  bounding box.north west)!#4!(path picture bounding
			box.north east)$);
			\fill [#2]
			($(path picture bounding box.south west)!#4!(path picture bounding
			box.south east)$)
			rectangle
			($(path picture  bounding box.north west)!#5!(path picture bounding
			box.north east)$);
			\fill [#3]
			($(path picture bounding box.south west)!#5!(path picture bounding
			box.south east)$)
			rectangle (path picture bounding box.north east);
}}}

\usepackage[many]{tcolorbox}
\tcbuselibrary{listings,skins}

\lstdefinestyle{mystyle}{
  xleftmargin=0pt,
   basicstyle={\footnotesize\ttfamily},
   aboveskip=3mm,
   belowskip=3mm,
   keywordstyle=\bfseries,
   showstringspaces=false,
  escapechar=?,
  language=Java
}
\definecolor{code_indent}{HTML}{CCCCCC}

\newtcblisting{mylisting}[2][]{
  arc=0pt, outer arc=0pt,
  listing only,
  listing style=mystyle,
  title={\large #2},
  #1
}

 \definecolor{dkgreen}{rgb}{0,0.6,0}
 \definecolor{gray}{rgb}{0.5,0.5,0.5}
 \definecolor{mauve}{rgb}{0.58,0,0.82}

\definecolor{cadmiumgreen}{rgb}{0.0, 0.42, 0.24}
\definecolor{verde}{rgb}{0.25,0.5,0.35}
\definecolor{jpurple}{rgb}{0.5,0,0.35}
\definecolor{darkgreen}{rgb}{0.0, 0.2, 0.13}

\usepackage[ruled,vlined,linesnumbered,lined]{algorithm2e}
\usepackage{algpseudocode}

\usepackage{mathtools,xparse}

\usepackage{tabu}
\usepackage{multirow}

\usepackage{pifont}
\usepackage{enumerate}
\usepackage[shortlabels]{enumitem}

\usepackage{pgfplotstable}
\usepackage{pgfplots}

\usepackage[noframe]{showframe}
\usepackage{framed}

 {\endMakeFramed}
 \definecolor{shadecolor}{gray}{0.85}

\definecolor{bgblue}{RGB}{245,243,253}
\definecolor{ttblue}{RGB}{91,194,224}

\mdfdefinestyle{mystyle}{%
  rightline=true,
  innerleftmargin=10,
  innerrightmargin=10,
  outerlinewidth=3pt,
  topline=false,
  rightline=false,
  bottomline=false,
  skipabove=\topsep,
  skipbelow=false
}

\newtcolorbox{myboxi}[1][]{
  breakable,
  title=#1,
  colback=white,
  colbacktitle=white,
  coltitle=black,
  fonttitle=\bfseries,
  bottomrule=0pt,
  toprule=0pt,
  leftrule=3pt,
  rightrule=3pt,
  titlerule=0pt,
  arc=0pt,
  outer arc=0pt,
  colframe=black!50,
}

\newtcolorbox{myboxii}[1][style=mystyle]{
  breakable,
  freelance,
  colback=white,
  colbacktitle=white,
  coltitle=black,
  fonttitle=\bfseries,
  bottomrule=0pt,
  boxrule=0pt,
  colframe=white,
  after skip=0pt,
  overlay unbroken and first={
    \draw[white!75!black,line width=3pt]
    ([yshift=-9pt]frame.north west) --
    ([yshift=9pt]frame.south west);
  },
}

\def\BibTeX{{\rm B\kern-.05em{\sc i\kern-.025em b}\kern-.08em
    T\kern-.1667em\lower.7ex\hbox{E}\kern-.125emX}}
    
\makeatletter
\newcommand{\linebreakand}{%
  \end{@IEEEauthorhalign}
  \hfill\mbox{}\par
  \mbox{}\hfill\begin{@IEEEauthorhalign}
}
\usepackage{balance}

\makeatother

\begin{document}

\title{Metamorphic Testing and Debugging \\of Tax Preparation Software}

\author{\IEEEauthorblockN{Saeid Tizpaz-Niari}
\IEEEauthorblockA{\textit{Computer Science Department} \\
\textit{University of Texas at El Paso}\\
saeid@utep.edu} \\
\IEEEauthorblockN{Shiva Darian}
\IEEEauthorblockA{\textit{Information Science Department} \\
\textit{University of Colorado Boulder}\\
shiva.darian@colorado.edu}
\and
\IEEEauthorblockN{Verya Monjezi}
\IEEEauthorblockA{\textit{Computer Science Department} \\
\textit{University of Texas at El Paso}\\
vmonjezi@miners.utep.edu} \\
\IEEEauthorblockN{Krystia Reed}
\IEEEauthorblockA{\textit{Psychology Department} \\
\textit{University of Texas at El Paso}\\
kmreed2@utep.edu}
\and
\IEEEauthorblockN{Morgan Wagner}
\IEEEauthorblockA{\textit{Psychology Department} \\
\textit{University of Texas at El Paso}\\
mrwagner@miners.utep.edu}
\\
\IEEEauthorblockN{Ashutosh Trivedi}
\IEEEauthorblockA{\textit{Computer Science Department} \\
\textit{University of Colorado Boulder}\\
ashutosh.trivedi@colorado.edu}
}

\IEEEtitleabstractindextext{
\begin{abstract} 
This paper presents a data-driven debugging framework to improve the trustworthiness of US tax preparation software systems. 
Given the legal implications of bugs in such software on its users,
ensuring compliance and trustworthiness of tax preparation software is of paramount importance.
The key barriers in developing debugging aids for tax preparation systems
are the unavailability of explicit specifications and the difficulty of obtaining oracles.
We posit that, since the US tax law adheres to the legal doctrine of precedent,
the specifications about the outcome of tax preparation software for an individual
taxpayer must be viewed in comparison with individuals that are deemed similar.
Consequently, these specifications are
naturally available as properties on the software requiring similar inputs provide
similar outputs. Inspired by the metamorphic testing paradigm, we dub these relations \emph{metamorphic relations} as they relate to structurally modified inputs. 

In collaboration with legal and tax experts, we explicated metamorphic relations for a set of challenging properties from various US Internal Revenue Services (IRS) publications including Form 1040 (U.S. Individual Income Tax Return), Publication 596 (Earned Income Tax Credit), Schedule 8812 (Qualifying Children and Other Dependents), and Form 8863 (Education Credits). 
While we focus on an open-source tax preparation software for our case study, the proposed framework can be readily extended to other commercial software. 
We develop a randomized test-case generation strategy to systematically validate the correctness of tax preparation software guided by metamorphic relations. We further aid this test-case generation by visually explaining the behavior of software on suspicious instances using easy-to-interpret decision-tree models. 
Our tool uncovered several accountability bugs with varying severity ranging from non-robust behavior in corner-cases (unreliable behavior when tax returns are close to zero) to missing eligibility conditions in the updated versions of software.
\end{abstract}
}


\maketitle
\IEEEdisplaynontitleabstractindextext

\section{General Abstract}
\label{sec:general-abstract}
The authors present a framework for supporting software developers and regulators in testing and auditing sociolegal critical software systems. 
The key barriers to the automation of the analysis for such software include: 1) the unavailability of correctness requirements and 2) the difficulty of obtaining ground truths for program inputs. 
The authors posit that, since the US legal system adheres to \emph{stare decisis} (to stand by things decided) doctrine, the specifications about the outcome of legal-critical software for an individual must be viewed in comparison with individuals that are deemed similar.
Such specifications are dubbed \emph{metamorphic relations} in this work due to their semblance with metamorphic testing paradigm from software engineering. 
The paper proposes a specification language to express metamorphic relations and uses it to automate the search for suspicious or incorrect behavior of a given software system and uses machine learning algorithms to present a succinct explanation of the identified vulnerabilities.

As a paradigmatic example of sociolegal critical software, the paper focuses on an open-source implementation of the US tax preparation software. 
The authors explicated metamorphic relations for a set of challenging properties from various US Internal Revenue Services (IRS) publications
including Form 1040 (U.S. Individual Income Tax Return), Publication 596 (Earned Income Credit), Schedule 8812 (Qualifying Children and Other Dependents), and Form 8863 (Education Credits). 
The framework is demonstrated to be useful in uncovering  several accountability bugs such as non-robust behavior in corner-cases and missing eligibility conditions.

\section{Introduction}
\label{sec:intro}
The ever-increasing complexity of income tax laws in the United States has rendered manual preparation of tax returns cumbersome and error-prone. According to the IRS, $90$ percent of tax filers filed taxes electronically in 2020~\cite{IRS-efile}.
Consequently, US tax preparation has grown into a $\$11.2$bn industry requiring the services of professional tax accountants or commercial tax preparation software. 
The use of software is increasing, and in 2020, over 72 million people prepared their taxes without the help of tax professionals, a 24 percent increase from 2019~\cite{IRS-rates}. 

Even though there are some freely available open-source alternatives~\cite{openTaxSolver,Ustaxes,Openfisca},
such tax preparation software and services are provided on ``AS-IS'' bases and may not go through rigorous software development process. 
The impact of bugs in software are aggravated by the fact that, in the US tax courts, individuals are accountable 
for any errors resulting from software bugs in such packages:
\begin{quote}
\emph{{\bf Langley v. Comm’r, T.C. Memo. 2013-22.} The misuse of tax preparation software, even if unintentional or accidental, is no defense to accuracy-related penalties under section 6662.} 
\end{quote}
However, users might not be aware of errors in the software.
And few checks currently exist to ensure the correctness of tax software~\cite{Regulating-Returns}. 
To protect consumers in the this process, we develop validation and debugging aids for open-source tax preparation software. 
Particularly because this problem has the potential to compound financial stresses
of the lower income earners, who are more likely to use freely available and unregulated software~\cite{Regulating-Returns}.

This paper presents a data-driven debugging framework to discover and explain bugs in tax preparation software systems. 
There are three concrete obstacles to this framework. 
\begin{itemize}
    \item \textit{Absence of Oracle.} The class of correctness requirements for tax preparation systems
    are not explicitly available since the correct tax-filing is highly subjective to individual taxpayers (this is known as \textit{oracle} problem~\cite{6963470}).
    \item \textit{Lack of trustworthy dataset.} While, in the absence of such explicit specification,
    one can recourse to data-driven approaches, it is difficult to obtain access to a valid dataset
    due to obvious privacy and legal concerns. 
    \item \textit{Computational infeasibility.}  Even when one can justify fabricating such a dataset,
    the final challenge is computational in nature: detecting and explaining the bug
    by comparing an individual decision to others over a large dataset based on a given similarity
    measure is computationally intractable.
\end{itemize}

Metamorphic testing~\cite{chen-original} is a software testing paradigm that tackles the oracle problem by considering software properties where the correctness of the software on an input does not require knowing the ``ground truth'' for that input; rather, the correctness can be validated by comparing the output of software for that input with the output of a slightly \emph{metamorphosed} one.  
For instance, consider a program implementing a search engine: while there is no way to verify that the results returned for a particular keyword are correct (oracle problem), it is reasonable to expect that a correct search engine should return fewer results for a more restricted keyword.

This paper makes a case for the suitability of \emph{metamorphic specifications} in testing and debugging tax preparation software (and by extension for legal software systems).  
We address aforementioned challenges by integrating metamorphic specifications with data-driven debugging in the following manner:

\begin{itemize}
\item \textit{Metamorphic Specifications.} The first obstacle in developing the debugging framework  is to explicate an appropriate notion of correctness requirements. 
Note that, given the relevant information about an individual, resolving the correct decision for
that individual requires accounting and legal expertise.
Hence, obtaining the oracle for testing and debugging purposes is impractically expensive.
Fortuitously, since the US tax law is a law code, it adheres to principle of \emph{common law} and implements \emph{stare decisis}, i.e. the legal doctrine of precedent: similar cases must follow similar rulings. 
As a corollary, the correctness of tax preparation software must
also be viewed in comparison with similar cases. 
In other words, the correctness properties can be expressed as relations (1) between two individual taxpayers with similar situations, i.e., a notion of horizontal equity in taxation~\cite{musgrave1990horizontal} and (2) between two taxpayers in different tax income buckets, i.e., a notion of vertical equity in taxation~\cite{10.1145/3531146.3533204}.   
Following the metamorphic testing~\cite{chen-original}, we call such properties \emph{metamorphic specifications}.
In collaboration with tax experts, we reviewed some critical correctness properties of tax preparation outcomes and observed that they are naturally expressible as metamorphic specifications. 
One key contribution of this paper is to explicate formal representations of these properties from the latest Internal Revenue System (IRS) documents. 

\item \textit{Test-case Generation.}  While it is an arduous task to compare an individual situation with the precedents, the availability of software artifacts implementing the tax laws permit us to query the tax outcomes for fictitious individuals similar to a given individual. 
This combined with the metamorphic nature of the requirements allows us validate the software system by comparing the results for a given (source) individual with its (metamorphosed) follow-ups.
We develop a random search strategy to sample source and follow-up test cases from the
metamorphic relations and label them as passed or failed.

\item \textit{Data-Driven Debugging.}
Given a set of test cases with `pass' or `fail' labels over a metamorphic relation,
decision trees, as a white-box ML model, are a natural choice
to discriminate between the classes and synthesize circumstances under which the
software fails~\cite{tizpaz2018differential,DBLP:conf/issta/Tizpaz-NiariC020}.
\end{itemize}

The proposed framework is implemented as the tool \toolname
(named after the US individual tax return form 1040).
We performed experiments on
\textit{OpenTaxSolver}~\cite{openTaxSolver} (tax years of 2018-2021),
a popular open-source tax preparation software~\cite{reddit-opentaxsolver,opensource-opentaxsolver},
in five aspects of disability, credits, and deductions that are known to be challenging and
error-prone~\cite{IRS-common-mistakes}. 
In particular, we explicated $16$ metamorphic
relations for the tax year of 2020 in those areas and with little effort, we managed to adapt them for the other tax years (2018, 2019, and 2021). We use these relations to generate test cases as well as oracles. Specifically,
our technique generates tens of thousands of random test cases using the metamorphic
properties. We explain the circumstances under which the software has both failed and passed cases
using CART decision tree algorithm~\cite{Breiman/1984/CART}. We provide debugging supports
in both spaces of input variables (fields/items in the tax preparation software)
and internal variables (conditions, loops, and function calls).
As a result, \toolname revealed three types of failures in \textit{OpenTaxSolver}:
missing some eligibility conditions (e.g., married people filing separately status is not eligible to
take earned income credits); software fails to satisfy the correctness requirements when the
computed tax returns get very close to zero (small non-zero values); and the updated software
that allows users to explicitly opt for an option does not satisfy some correctness requirements
in the corner cases.

\section{Background} 
\label{sec:background}

\begin{figure*}[!htb]
    \centering
    \includegraphics[width=1.0\textwidth]{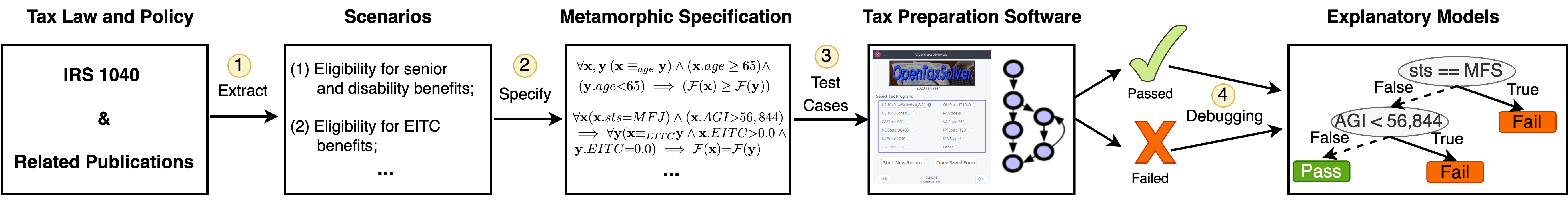}
    \caption{
    \textbf{General Framework using Disability and EITC benefits.} Our approach specifies metamorphic properties
    from relevant tax policies. Then, it generates random test cases and infers decision trees
    to localize circumstances under which the software fails to satisfy metamorphic requirements.}
    \label{fig:overall-framewok}
\end{figure*}

The US tax code is based on over 2,600 pages of legal statutes from congress.
These statutes are accompanied by IRS regulations, rulings, and clarifications,
together amounting to nearly 9,000 pages.
Lawsky et al.~\cite{formalizingCode,DBLP:conf/icail/PertierraLHO17} have extensively studied the structure of the US tax code and have made a convincing case for the need of formal legal reasoning. 
The structure has gone so complex that the IRS publishes updated instructional resources every fiscal year. 
Lawsky attributes part of the complexity of US tax codes to \emph{internal dependencies},
referring to the interlinked nature of calculations across sections of tax forms ~\cite{formalizingCode}. This is evident as 97\%\ of references in the Internal Revenue Code (IRC) are internal. Internal dependencies can lead to \emph{legal avoidance}, when the tax code is not adhered to as intended ~\cite{DBLP:conf/icail/PertierraLHO17} and can occur when the IRC is misinterpreted.
At this point, there has been limited effort to test U.S. tax software that attempts to prepare tax filings in a manner consistent with the extensive tax code.

A study by Nina E. Olson
(National Taxpayer Advocate), published by IRS~\cite{testimony08},
analyzed several (free) tax filing software for the tax year of 2005. The study designed four fictitious scenarios, including some in which the taxpayer qualified for special tax benefits
enacted to help Hurricane Katrina victims (See~\cite[p.11-p.19]{testimony08}). 
The study found that several of the software did not prompt the taxpayers about these provisions.
Therefore, taxpayers ended up not getting the benefits designed for them. 
Despite being hypothetical, the approach uncovered flaws in the  tax filing software and underscored the need for trustworthiness besides software privacy and security. 
Our approach seeks to automate such discoveries
in tax preparation software and provide intuitive explanations.

The absence of logical specifications and oracle in the context of tax preparation software, led us to investigate metamorphic testing. 
Metamorphic testing~\cite{chen-original} is primarily developed to counter the lack of oracle~\cite{6963470} (black box module to decide whether the output of the system is correct for a given input). 
The idea behind metamorphic testing is to establish correctness by reasoning
about relations between multiple input-output behaviors~\cite{metamorphic}. 
One example is to validate search engines. While we do not know the ground truth
number of return items for any search query $x$, 
we can expect that for any query $x_1 = x$ and its restrictive one $x_2 = x + k$
with more keywords, the number of retrieved items for $x_1$ is greater than $x_2$.
For another example, consider
a program that implements \texttt{sin}(x) function. While we do not know
the exact outcome of function for an arbitrary $x$, we can establish that for any
$x_1$ and its metamorphosed $x_2 = 2*\pi + x_1$, \texttt{sin}($x_1$) = \texttt{sin}($x_2$).
In this example, $x_1$ is a source (base) test, and $x_2$ is a follow-up (metamorphosed) test constructed by transforming the source to satisfy the metamorphic relation
between $x_1$ and $x_2$ and check the equality between outcomes.

Metamorphic properties have been successfully used to validate a variety of software systems.
They have been adapted to validate machine learning
classifiers~\cite{MetamorphicMLClassifiers}, autonomous vehicles~\cite{zhou2019a},
Google/Yahoo search engine~\cite{zhou2012automated}, and social media
apps like Facebook~\cite{ahlgren2021testing}. To the best of our knowledge,
this work is the first to adapt metamorphic relations in
\textit{tax preparation software} domain as a sociotechnological system.

\section{Overview}
\label{sec:overview}
As a demonstration of our framework (shown in Figure~\ref{fig:overall-framewok}),
this section describes how we tested OpenTaxSolver~\cite{openTaxSolver} (versions
for the tax years 2018-2021) in one of $5$ domains considered in this paper: the Earned Income Tax Credit ($EITC$). 
The earned income tax credit is designed to lower the tax burden
for low and moderate income workers.

\vspace{0.5em}
\noindent\textbf{Extract Scenarios.} The first step involves carefully examining the rules and
regulations. For the $EITC$, we extract scenarios from
Worksheet 1 in Publication 596~\cite{pub596}.

\vspace{0.5em}
\noindent\textbf{Metamorphic properties.} 
We specify $4$ metamorphic relations for $EITC$,
as shown in Table~\ref{tab:metamorphic-relations} with EITC (3)-(6) for the tax year 2020,
and update these properties for the tax years 2018, 2019, and 2021 as shown in Table~\ref{tab:metamorphic-relations-changes}.
Property \#3 specifies that married filing separately ($MFS$) is not eligible to take $EITC$ credits in the tax year 2020,
noting that the same condition applies for the tax years 2018 and 2019, but the requirement has changed for the tax year 2021
where $MFS$ status is eligible to take the $EITC$ credits (see Table~\ref{tab:metamorphic-relations-changes}).
Property \#4 describes that individuals who are married filing jointly ($MFJ$) are eligible
if their adjusted gross income ($AGI$) is below or equal to $56,844$ for the tax year 2020. Note that
different years have different thresholds of eligibility as shown in Table~\ref{tab:metamorphic-relations-changes}.
Property \#5 says that an eligible individual with $EITC$ claim (Line $27$ of
IRS $1040$) should receive equal or higher tax returns compared to metamorphosed individuals who are not
eligible due to their $AGI$, conditions on their dependents, or zero claims of $EITC$.
Finally, property \#6 establishes that among two similarly eligible individuals,
the one with the higher $EITC$ claims should receive a higher tax return. Other metamorphic
relations are described and formulated in Section~\ref{sec:experiments}-RQ1.

\vspace{0.5em}
\noindent\textbf{Random Test-Case Generations.} 
We generate test cases from these metamorphic properties. 
We encode each metamorphic property
as a search problem where the objective is to find an individual and its
metamorphosed one such that their deviants from expected outcomes are maximum.
We sample data points uniformly at random to generate the source test case
and perturb it to generate follow-ups from its neighbors. 
We sketch the detail of search algorithm to generate test cases in
Algorithm~\ref{alg:random-testing}. 
The results of experiments for generating
test-cases are shown in Table~\ref{table:experiments}.
Within $15$ minutes of search,
our implementation of the algorithm in \toolname generate over $19$k,
$18$k, $36$k, and $36$k, for properties \#3 to \#6, respectively, for the 2020
version (see Table~\ref{table:experiments} for the complete experimental
results). The results show that the software fails for some or all test cases
of properties \#3-\#4 in the version 2020, but passes for all test cases of
these two properties in the versions 2018 and 2019. While, the 2021 version 
of software fails for property \#3, it passes all test cases for property \#4.
We found that the updated tax policy where the $MFS$ status is eligible for $EITC$
in 2021 leads to the expected behavior of software for property \#4 without explicit
modifications compared to the faulty 2020 version! The details of experiments for test-case generations
can be found in Section~\ref{sec:experiments}-RQ2.

\vspace{0.5em}
\noindent\textbf{Explaining Failure Circumstances.} Given the set of passed and failed inputs,
we use a decision tree algorithm (detailed in Section~\ref{sec:approach})
to infer conditions under which software fails to satisfy the metamorphic relations.
Since the software (ver. 2020) fails on all test cases for property \#4,
the premises of metamorphic relation itself simply explain the failure circumstances: the tax
software does not appropriately check eligibility on the basis of $AGI$ for $EITC$ credits.
For property \#3, we also found that the software fails in most cases, but
it passes for a few corner cases. We use the decision tree to find an explanation
for this phenomenon (see Table~\ref{tab:debugging}, the 1st row). It shows that
the software satisfies the metamorphic relation when $AGI$ is below \$16,720
and the federal tax return is below \$80. This might be expected due to eligibility
for some credits when the tax returns are very low.
The decision tree artifacts for other properties are also shown in Table~\ref{tab:debugging}.
Beyond $EITC$, the decision trees show that software might fail due to the numerical
approximation for small non-zero values and missing to consider all conditions when updated
to allow to opt for an option.
The experiments for debugging are provided in Section~\ref{sec:experiments}-RQ3.

\section{Problem Statement}
\label{sec:problem}

The \emph{functional} model of tax software consists of a tuple $(X, \mathcal{F})$ where
$X = \set{X_1, X_2, \ldots, X_n}$ is the set of variables corresponding to various fields
about an individual in the tax return form and $\mathcal{F}: \mathcal{D}_1 \times \mathcal{D}_2 \times \cdots \times \mathcal{D}_n \to \Real_{\geq 0}$ is the \emph{federal tax return} computed by the software,
where $\mathcal{D}_i$ is the domain of variable $X_i$.
We write $\mathcal{D}$ for $\mathcal{D}_1 \times \mathcal{D}_2 \times \cdots \times \mathcal{D}_n$.

These variables correspond to intuitive labels such as ${\tt age}$ (numerical variable),
${\tt blind}$ (Boolean variable), and ${\tt sts}$ (filing status with values such as MFJ, married filing jointly, and MFS, married filing separately). 
For an individual ${\bf x} \in \Dd$, we write ${\bf x}(i)$ for the value of $i$-th variable,
or ${\bf x}.{\tt lab}$ for the value of variable ${\tt lab}$.
Let $\mathcal{L}$ be the set of all labels.

For labels $L \subseteq \mathcal{L}$ and inputs ${\bf x} \in \mathcal{D}$ and ${\bf y} \in \mathcal{D}$, we say that ${\bf y}$ is a metamorphose of ${\bf x}$ with the exceptions of labels $L$, and we write ${\bf x} \equiv_{L} {\bf y}$ if 
$\forall \ell \not \in L \text{ we have that } {\bf x}.\ell = {\bf y}.\ell$.
A metamorphic relation is a first-order logic formula with variables in $X$, constants from domains in $\mathcal{D}$, relation $\equiv_L$, comparisons $\set{<, \leq, = , \geq, >}$  over numeric variables, predicate $\neg$ (negation) for Boolean valued labels, real-valued function for federal tax return $\mathcal{F}: \Dd \to \Real$, Boolean connectives $\wedge$, $\lor$, $\neg$, $\implies$, $\iff$, and quantifiers $\exists x. \phi(x)$ and $\forall x. \phi(x)$ with natural interpretations. 
W.l.o.g., we assume that the formulas are given in the prenex normal form, i.e. a block of quantifiers followed by a quantifier free formula.
Using metamorphic relations, we can express the property,

\begin{quote}
    for two individuals that differ only in age, the federal tax return of older individual must be greater than or equal to that of the younger one,
\end{quote}
 using the following metamorphic property $\phi$:
 \[
 \forall {\bf x}, {\bf y} \: ({\bf x} \equiv_{age} {\bf y}) \wedge ({\bf x}.age \geq {\bf y}.age) \implies (\mathcal{F}({\bf x}) \geq \mathcal{F}({\bf y})).
 \]

\vspace{0.5em}
\noindent\textbf{Falsification Problem.} Given a tax preparation software $(X, \mathcal{F})$ and a metamorphic property $\phi$, the falsification problem is to explore the input space of the software to discover inputs that falsify the property. 
Metamorphic properties containing only one variable are classical properties relating inputs to outputs; however for properties with more than one quantified variables, multiple inputs are required to evaluate the property. 
Observe that the metamorphic property falsification problem is PSPACE-hard as it is expressive enough to encode True Quantified Boolean formula (TQBF) problem (when the input variables are Boolean and program logic encode Boolean constraints). 
Hence, it is intractable to provide an exhaustive search procedure. To overcome this challenge, we propose randomized search-based testing approach. 

Depending upon the number and nature of of quantifiers, multiple inputs need to be sampled to ascertain the status of the property. 
Moreover, when the input space is large or infinite, we may not be able to falsify subformulae of the form $\exists {\bf x}. \phi({\bf x})$ as it would require testing the whole state space. To overcome this challenge in practice, we recourse to statistical testing guided by randomized exploration of the state space. 
In the next section, we present a sampling strategy to falsify properties of the form $\forall x. \phi(x)$;
this procedure can be recursively invoked to provide the procedure for metamorphic properties
with arbitrary quantification.


\section{Approach}
\label{sec:approach}

\begin{algorithm}[t!]
{
	\DontPrintSemicolon
	\KwIn{Tax preparation software $\Pp$,
	initial input seeds $I$,
	metamorphic property $p$, a tolerance threshold $\delta$,
	a Bayes factor $B$, a lower-bound on the confidence $\theta$,
	and timeout $T$.}
	\KwOut{Passed/Failed, test cases, decision tree}

	$(x_{m}, \Delta_{FTR}, res)$ $\gets$ \textsc{Sample}($I$),~0,~${\tt True}$

    \While{\texttt{time}() - $start\_time$ $<$ T}
    {
        $k$ $\gets$ 0
        
    	$x_1$ $\gets$ \textsc{UniformPerturb}($x_{m}$, $p$)

    	$x_2$ $\gets$ \textsc{UniformPerturb}($x_1$, $p$)
    	
    	$\Delta$ $\gets$ \textsc{Distance}($\Pp(x_1)$, $\Pp(x_2)$)
    	
    	\If{$\Delta$ $>$ $\delta$}{
        	$I$.\textsc{Add}\big(($x_1$,$x_2$),`\textit{failed}'\big)
        	
        	\If{$\Delta$ $>$ $\Delta_{FTR}$}{
        	     $x_{m}$ $\gets$ $x_1$
        	     
        	     $\Delta$ $\gets$ $\Delta_{FTR}$ 
        	}

    	    $res$ $\gets$ ${\tt False}$
    	}
    	\Else{
    	    $I$.\textsc{Add}\big(($x_1$,$x_2$),`\textit{passed}'\big)  
    	    
    	    $k$ $\gets$ $k$ + $1$

    	    \If{$k$ $<$ $\frac{-\log B}{\log \theta}$}
    	    {
    	        Go to $5$
    	    }
    	}
	}
	\If{$I[$`passed'$]$ = \{\} $\lor$ $I[$`failed'$]$ = \{\}}
	{
	    \Return $p$
	}
	\Else{
            $I'$ $\gets$ \textsc{FeatureEngineering}($I$)
            
	    $t$ $\gets$ \textsc{DTClassifier}($I'$)
	}

	\Return $res$, $I$, $t$.
	\caption{\textsc{RandomTestCaseGeneration}}
	\label{alg:random-testing}
}
\end{algorithm}

We develop an efficient metamorphic testing and debugging to uncover
bugs in tax preparation software. 
Algorithm~\ref{alg:random-testing} sketches different steps in our approach.
We start from some initial seed inputs $I$. In each step,
the search strategy selects and perturbs inputs from the set of
promising inputs (those maximizing the deviant from the expected outcome) and generates
a base test case (line $4$). Then, it perturbs the base to generate follow-up cases (line $5$).
Next, it executes the base and follow-up inputs on the target software and quantifies
the deviant between the outcomes (line $6$). If the deviant is more than a threshold, it
adds the inputs with `failed' label (line $8$). Furthermore, if the deviant is higher than
any previous ones, it adds the inputs as promising ones (line $9$). Otherwise, it adds the 
inputs with `passed' label (line $12$). However, the absence of evidence for failures
does not mean correctness. This is indeed an important shortcoming for metamorphic-based random
test generations. Our algorithm instead uses a statistical hypothesis testing to provide statistical confidence on the
absence of failures, starting from the base test case. The null and alternative hypotheses are
\[
\Hh_0: \pP(x_1) \geq \theta,~~\Hh_1:\pP(x_1) < \theta
\]
where $\pP(x_1)$ is the probability that follow-up test cases, starting from $x_1$ as the source test,
are passed, $\theta$ is the lower-bound on the probability, $\Hh_0$ is null hypothesis, and
$\Hh_1$ is the alternative. Our goal is to witness enough passed test cases to accept $\Hh_0$
as opposed to $\Hh_1$. There are multiple ways to conduct such statistical testing. 
The sequential probability ratio test and a Bayes factor are examples. We follow
Jeffreys test~\cite{jha2009bayesian,sankaranarayanan2013static}, a variant of Bayes factor,
with a uniform prior to find a lower-bound on the number of
successive samples $K$ that sufficient for us to convince $\Hh_0$:
\[
K \geq \lceil(-\log_2 B)/(\log_2 \theta)\rceil  
\]
where $B$ in numerator is Bayes factor and can be set to $100$ for a very strong
evidence. For instance, to achieve a $\theta = 0.95$, we are required to
set $K \geq 90$ to be highly confident on accepting $\Hh_0$. Lines 13-17
show this hypothesis testing. 

After generating the test suite, our algorithm uses a data-driven approach to infer circumstances
under which the software fails. Given the set of test cases with `passed' and
`failed' labels, the problem appears to be a standard classification to discriminate between
the two labels~\cite{kampmann2020does}. However, the generated samples are not independent of the follow-up
samples are dependent on the base samples. To overcome
this challenge and enable standard decision trees to provide intuitive explanations,
we use a feature engineering technique to define two copies of features: base features
and follow-up features (line 21). For example, for feature $AGI$, 
we now have two features $AGI\_1$ and $AGI\_2$ for the evaluation of base and follow-up cases. 
Then, we combine the base and follow-up test cases to represent them
with a single test instance. In the $AGI$ example with $AGI$=10k for the base and $AGI$=12k for
the follow-up test cases, we convert them into one test case with $AGI\_1$=10k and $AGI\_2$=12k.
This feature engineering enables us to directly infer decision tree models with an off-the-shelf
algorithm (line 22). 

\section{Experiments}
\label{sec:experiments}

\begin{table*}[!t]
    \caption{Metamorphic properties for five domains in the US tax policies.
    $\mathcal{F}$ is federal tax return where negative
    values mean the individual owns payment to the IRS, $sts$ is filing status, $s\_lab$ is spouse's field $lab$,
    $MFJ$: married filing jointly, $MFS$ is married filing
    separately, $AGI$ is adjusted gross income, $L27$ is line $27$ of IRS $1040$ for Earned Income Tax Credit (EITC), $QC$ is the number of qualified children, $OD$ is the number of other dependents, $CTC$ is child tax credits,
    $L19$ is line $19$ of IRS $1040$ for Child Tax Credit ($CTC$),
    $L29$ is line $29$ of IRS $1040$ for Education Tax Credit ($ETC$), $MDE$ is medical/dental expenses reported in line $1$ of schedule $A$, $iz$ is to use itemized deductions ($ID$) vs. standard deductions, and $L12$ is total itemized deductions ($ID$) from schedule $A$.}
    \label{tab:metamorphic-relations}
    \centering
    \begin{tabular}{ |p{0.5cm}|p{1.3cm}|p{15cm}|  }
     \hline
     Id & Domain & Metamorphic Property \\
     \hline
    1  & Disability & $\forall {\bf x}, {\bf y} (({\bf x} {\equiv_{age}} {\bf y}) \wedge ({\bf x}.age {\geq} 65) \wedge ({\bf y}.age {<} 65 ))
    \lor (({\bf x} {\equiv_{blind}} {\bf y}) \wedge ({\bf x}.blind \wedge \neg {\bf y}.blind))
    \implies \mathcal{F}({\bf x}) \geq \mathcal{F}({\bf y})$ \\
     \hline
     2  & Disability &  $\forall {\bf x} ({\bf x}.sts = MFJ) \implies \forall {\bf y} (({\bf x} \equiv_{s\_age} {\bf y}) \wedge ({\bf x}.s\_age \geq 65) \wedge ({\bf y}.s\_age < 65 ))
    \lor (({\bf x} \equiv_{s\_blind} {\bf y}) \wedge ({\bf x}.s\_blind \wedge \neg {\bf y}.s\_blind))
    \implies \mathcal{F}({\bf x}) \geq \mathcal{F}({\bf y})$ \\
    \hline
     3  & EITC  & $\forall {\bf x} ({\bf x}.sts = MFS) \implies \forall {\bf y} ({\bf x} {\equiv_{L27}} {\bf y} \wedge {\bf x}.L27 >0.0 \wedge {\bf y}.L27 = 0.0) \implies \mathcal{F}({\bf x}){=}\mathcal{F}({\bf y})$ \\
     \hline
     4  & EITC  & $\forall {\bf x} ({\bf x}.sts{=}MFJ) \wedge ({\bf x}.AGI{>}56,844) \implies \forall {\bf y} ({\bf x} {\equiv_{L27}} {\bf y} \wedge {\bf x}.L27{>}0.0 \wedge {\bf y}.L27{=}0.0){\implies}\mathcal{F}({\bf x}){=} \mathcal{F}({\bf y})$ \\
    \hline
     5  & EITC & $\forall {\bf x} ({\bf x}.sts{=}MFJ){\implies}\forall {\bf y} ({\bf x} {\equiv_{AGI}} {\bf y} \wedge {\bf x}.AGI{\leq}56,844 \wedge {\bf y}.AGI{>}56,844) \lor ({\bf x} {\equiv_{L27}} {\bf y} \wedge {\bf x}.L27{>}0.0 \wedge {\bf y}.L27{=}0.0) \lor ({\bf x} {\equiv_{QC}} {\bf y} \wedge {\bf x}.QC{\geq}{\bf y}.QC){\implies}\mathcal{F}({\bf x}){\geq} \mathcal{F}({\bf y})$ \\
     \hline
     6  & EITC & $\forall {\bf x} ({\bf x}.sts{=}MFJ){\wedge}({\bf x}.AGI{\leq}56,844)
     {\implies}\forall {\bf y} (({\bf x} {\equiv_{L27}} {\bf y}){\wedge}{\bf x}.L27{\geq}{\bf y}.L27){\implies}\mathcal{F}({\bf x}){\geq}\mathcal{F}({\bf y})$ \\ 
     \hline     
     7  & CTC & $\forall {\bf x} ({\bf x}.sts{=}MFS){\wedge}({\bf x}.AGI{\leq}200k)\forall{\bf y} (({\bf x} {\equiv_{L19}} {\bf y}){\wedge}({\bf x}.L19{\geq}{\bf y}.L19)){\implies}\mathcal{F}({\bf x}){\geq}\mathcal{F}({\bf y})$)\\   
     \hline 
     8  & CTC & $\forall {\bf x}, {\bf x'} ({\bf x}.sts {=} {\bf x'}.sts {=} MFJ) 
     {\wedge}({\bf x}.AGI {<} 400k){\wedge}({\bf x'}.AGI {\geq} 400k) 
     {\wedge}{\lceil}{\bf x'}.AGI{-}400k{\rceil}_{1k}*0.05{<} {\bf x'}.QC*2k+{\bf x}.OD*0.5k 
     {\implies} 
     \forall {\bf y}, {\bf y'} ({\bf x} {\equiv_{\{QC,OD\}}} {\bf y}){\wedge}({\bf x'} {\equiv_{\{QC,OD\}}} {\bf y'})  
     \wedge  (0 {\leq} {\bf y}.QC {=} {\bf y'}.QC {\leq} {\bf x}.QC {=} {\bf x'}.QC \leq 10)
     \wedge  (0 {\leq} {\bf y}.OD {=} {\bf y'}.OD \leq {\bf x}.OD {=} {\bf x'}.OD \leq 10)
     \implies (\mathcal{F}({\bf x}) - \mathcal{F}({y})) \geq (\mathcal{F}(x') - \mathcal{F}(y'))$\\       
     \hline
     9  & ETC  & $\forall {\bf x} ({\bf x}.sts = MFS) \implies \forall {\bf y} ({\bf x} {\equiv_{L29}} {\bf y} \wedge {\bf x}.L29 >0.0 \wedge {\bf y}.L29 = 0.0) \implies \mathcal{F}({\bf x}){=}\mathcal{F}({\bf y})$ \\
     \hline
     10 & ETC  & $\forall {\bf x} ({\bf x}.sts{=}MFJ) \wedge ({\bf x}.AGI{\geq}180k) \implies \forall {\bf y} ({\bf x} {\equiv_{L29}} {\bf y} \wedge {\bf x}.L29{>}0.0 \wedge {\bf y}.L29{=}0.0){\implies}\mathcal{F}({\bf x}){=} \mathcal{F}({\bf y})$ \\
    \hline
     11 & ETC  & $\forall {\bf x} ({\bf x}.sts{=}MFJ) \wedge ({\bf x}.AGI{\leq}160k) \implies \forall {\bf y} ({\bf x} {\equiv_{L29}} {\bf y} \wedge {\bf x}.L29{\geq}{\bf y}.L29){\implies}\mathcal{F}({\bf x}){\geq} \mathcal{F}({\bf y})$ \\
    \hline
     12  & ETC & $\forall {\bf x}, {\bf x'} ({\bf x}.sts {=} {\bf x'}.sts {=} MFJ) 
     \wedge ({\bf x}.AGI {\leq} 160k) \wedge (160k{<}{\bf x'}.AGI {<} 180k) 
     \implies
     \forall {\bf y}, {\bf y'} (({\bf x} \equiv_{L29} {\bf y}) \wedge ({\bf x'} \equiv_{L29} {\bf y'}) \wedge ({\bf x}.L29={\bf x'}.L29 \geq {\bf y}.L29={\bf y'}.L29)) 
     {\implies}(\mathcal{F}({\bf x}) - \mathcal{F}({y})) \geq (\mathcal{F}(x') - \mathcal{F}(y'))$\\       
     \hline
     13  & ID & $\forall {\bf x},{\bf y} ({\bf x} {\equiv_{MDE}} {\bf y}) \wedge ({\bf x}.MDE{\leq}{\bf x}.AGI*7.5\%)  \wedge ({\bf y}.MDE{=}0.0) \implies\mathcal{F}({\bf x}){=}\mathcal{F}({\bf y})$ \\
     \hline 
     14  & ID & $\forall {\bf x} (\neg {\bf x}.iz) \implies \forall {\bf y} ({\bf x} {\equiv_{MDE}} {\bf y} \wedge {\bf x}.MDE{>}0.0 \wedge{\bf y}.MDE{=}0.0) \implies\mathcal{F}({\bf x}){=}\mathcal{F}({\bf y})$ \\
     \hline
     15  & ID & $\forall {\bf x} ({\bf x}.sts{=}MFJ){\implies}\forall {\bf y} (({\bf x} {\equiv_{iz,L12}} {\bf y}){\wedge}({\bf x}.iz{\wedge}{\neg}{\bf y}.iz){\wedge}({\bf x}.L12{\leq}24.8k{\wedge}{\bf y}.L12{=}0.0)){\implies} \mathcal{F}({\bf x}){\leq}\mathcal{F}({\bf y})$ \\
     \hline     
     16  & ID & $\forall {\bf x} ({\bf x}.sts{=}MFJ){\implies}\forall {\bf y} (({\bf x} {\equiv_{iz,L12}} {\bf y}){\wedge}({\bf x}.iz{\wedge}{\neg}{\bf y}.iz){\wedge}({\bf x}.L12{>}24.8k{\wedge}{\bf y}.L12{=}0.0)){\implies} \mathcal{F}({\bf x}){\geq}\mathcal{F}({\bf y})$ \\
     \hline     
     \end{tabular}
\end{table*}

\vspace{0.5em}
\noindent\textbf{Implementations.}
We implement test-case generations in Python using the XML parser library to define the tax field variables and their domains. We instrument the tax preparation software using \texttt{llvm-cov} to collect code coverage information for each input trace (e.g., execution count of each line of code). We implement the decision trees in scikit-learn framework~\cite{scikit-learn} using the CART tree library.

\vspace{0.5em}
\noindent \textbf{Environment and Setup.}
We run all the experiments on an Ubuntu 20.04.4 LTS OS  sever with AMD Ryzen Threadripper PRO 3955WX 3.9GHz 16-cores X 32 CPU and two NVIDIA GeForce RTX 3090 GPUs.
In Algorithm~\ref{alg:random-testing}, we set $B$ to 100, $\theta$ to 0.95, $\delta$ to 1.0, and $T$ to 900.  

\vspace{0.5em}
\noindent \textbf{Open-source Tax Preparation Software.}
We used four different versions of \textit{OpenTaxSolver}~\cite{openTaxSolver} for the tax year 2018 to 2021.
The software includes the US individual federal tax return as well as the tax return computations for multiple US states (e.g., CA, MA, and NC).
The functionalities used for computing the federal tax return include 6,017 lines of code for the tax year 2020.
The number of input fields for the tax year of 2018, 2019, 2020, and 2021 are 92, 100, 110, and, 187, respectively. Many of these fields
are real numbers, but a few fields are binary and categorical.  

\vspace{0.5em}
\noindent \textbf{Research Questions.}
We seek to answer the following three questions using our tool \toolname in the defined setup.

\begin{itemize}
\item \textbf{RQ1.} Are metamorphic relations useful to capture the legal requirements of tax preparation software?

\item \textbf{RQ2.} Can randomized algorithm with Bayesian guarantees be effective in testing tax preparation software against the metamorphic specifications?

\item \textbf{RQ3.} Could fault localization techniques help pinpoint the root of failures in the internal and input spaces?
\end{itemize}

\begin{tcolorbox}[boxrule=1pt,left=1pt,right=1pt,top=1pt,bottom=1pt]
The source code and implementation of \toolname is available at: \url{https://github.com/Tizpaz/TenForty}
\end{tcolorbox}

\subsection{Metamorphic Relations (RQ1)}
\noindent\textbf{Tax Policies.} We consider aspects of the U.S. Individual Income
Tax Return that relate to disability, credits, and deductions. These domains 
are known to be notoriously challenging~\cite{IRS-common-mistakes}. We focus on
fields related to the standard deductions for senior and disable individuals; Earned Income Tax Credit (EITC)~\cite{pub596}, a refundable tax credits for lower-income households; Child Tax Credit (CTC), a nonrefundable credits to reduce the taxes owed based on the number of qualifying children under the age of 17~\cite{8812};
Educational Tax Credit (ETC) that help students with the cost of higher education by lowering
their owed taxes or increasing their refund~\cite{8863};
and Itemized Deduction (ID) that is an option for taxpayers with significant tax deductible expenses~\cite{1040sa}.

\vspace{0.5em}
\noindent\textbf{Metamorphic Relations.} We use scenarios and examples described in these policies to synthesize metamorphic relations. Table~\ref{tab:metamorphic-relations} shows $16$ metamorphic relations in $5$ domains for the tax year 2020. For properties
\#9 to \#12, we assume $MAGI$ (modified adjusted gross income) is equivalent to $AGI$~\cite{diff-AGI-MAGI}.
Next, we provide a brief explanation of some of these properties. 

\begin{itemize}
    \item 
\textit{Property \#1.} A senior (over age of 65) or bind individual must receive similar or better tax benefits when compared to a person without the
disability or seniority who is similar in every other aspect (due to higher standard deductions for seniors and blinds).

\item \textit{Property \#2.} An individual with the married filing jointly ($MFJ$) status
with a disabled/senior spouse must receive similar or higher tax benefits compared to a similar individual but without
the disabled or senior spouse.

\item \textit{Property \#3.} An individual with the married filing separately ($MFS$) status
is ineligible for EITC in 2020. 

\item \textit{Property \#4.} An individual with the married filing jointly ($MFJ$) status with 
$AGI$ over $56,844$ is ineligible for EITC in 2020.

\item \textit{Property \#5.} An individual who qualifies for EITC must receive a higher return than a similar unqualified one.

\item \textit{Property \#6.} Among two qualified
individuals with EITC, one with higher EITC claims receives higher or equal benefits.

\item \textit{Property \#7.} Among two qualified
married filing jointly ($MFJ$) individuals, one with higher child tax credits receives higher or equal  tax return benefits.

\item \textit{Property \#8.} 
This 4-property requires a comparison between
four ``similar'' individuals since there is a relation between two variables of interests:
$AGI$ and the number of qualified children/others to claim a CTC.
An individual with more qualified dependents
must receive higher or similar tax return benefits than an individual with fewer dependents after adjusting for the effects
of income levels on the calculations of both the final return and the amounts of CTC claims. Expressing this property requires
holding the income of two individuals the same per each qualified number of children/others.

\item \textit{Property \#9.} An individual with the married filing separately ($MFS$) status is 
ineligible for ETC in 2020.

\item \textit{Property \#10.}  An individual with the married filing jointly ($MFS$) status
with $AGI$ over $180k$ is ineligible for ETC.

\item \textit{Property \#11.} A qualified
individual with $AGI$ below $160k$ who claims ETC received higher or similar
tax return benefits compared to a similar individual who is ineligibility or
does not claim ETC for the tax year 2020.

\item \textit{Property \#12.} 
This 4-property requires a comparison between four
``similar'' individuals as the rule changes for individuals with $AGI$ below $160$k and between $160$k and $180$k.
By holding $AGI$ constant between two individuals with $AGI$ below $160$k (varying the $ETC$ claims) and 
two individuals with $AGI$ between $160$k and $180$k (varying
the $ETC$ claims with the same rate), the property requires that individuals with lower income (below $160k$) receive higher or similar
tax returns.

\item \textit{Property \#13.} An individual who files
with medical/dental expenses ($MDE$) below 7.5\% of their $AGI$ and itemizes their
deductions receives the same return as a similar individual with no $MDE$ claims.

\item \textit{Property \#14.} When filing with a standard deduction, total itemized deduction (Line $12$) must have no effect on the tax returns.

\item \textit{Property \#15.} An individual who files 
with itemized deductions below the standard deductions receive a lower or similar
tax return benefits compared to a similar individual who used the standard deductions.

\item \textit{Property \#16.} An individual who files 
with itemized deductions above the standard deductions receive a higher or similar
tax return benefits compared to a similar individual who claims standard deductions.
\end{itemize}

\begin{table}[!t]
    \caption{Updated metamorphic relations for tax years of 2018, 2019, and 2021 with respect to year 2020 (Table~\ref{tab:metamorphic-relations}).}
    \label{tab:metamorphic-relations-changes}
    \centering
    \begin{tabular}{ |p{0.6cm}|p{2.1cm}|p{2.1cm}| p{2.1cm} |}
     \hline
     Id & Year 2018 & Year 2019 & Year 2021 \\
     \hline
     1,2  & No Change & No Change & No Change \\
     \hline
     3  & No Change & No Change & $\mathcal{F}({\bf x}){\geq}\mathcal{F}({\bf y})$ \\
     \hline
     4  & ${\bf x}.AGI{>}54,884$ & ${\bf x}.AGI{>}55,952$ & ${\bf x}.AGI{>}57,414$ \\
     \hline
     5  & ${\bf x}.AGI{\leq}54,884 \wedge {\bf y}.AGI{>}54,884)$ & ${\bf x}.AGI{\leq}55,952 \wedge {\bf y}.AGI{>}55,952)$ & ${\bf x}.AGI{\leq}57,414 \wedge {\bf y}.AGI{>}57,414)$ \\
     \hline
     6  & ${\bf x}.AGI{\leq}54,884$ & ${\bf x}.AGI{\leq}55,952$ & ${\bf x}.AGI{\leq}57,414$ \\
     \hline 
     7-13  & No Change & No Change & No Change \\
     \hline 
     14  & Not Possible & Not Possible & No Change \\
     \hline 
     15  & ${\bf x}.L8{\leq}24.0k \implies \mathcal{F}({\bf x}){=}\mathcal{F}({\bf y})$ & ${\bf x}.L9{\leq}24.4k \implies \mathcal{F}({\bf x}){=}\mathcal{F}({\bf y})$ & ${\bf x}.L12{\leq}25.1k \implies \mathcal{F}({\bf x}){\leq}\mathcal{F}({\bf y})$  \\
     \hline 
     16  & ${\bf x}.L8{>}24.0k$ & ${\bf x}.L9{>}24.4k$ & ${\bf x}.L12{>}25.1k$ \\
     \hline 
     \end{tabular}
\end{table}

\vspace{0.5em}
\noindent\textbf{Metamorphic Relations in Different Years.} As the tax law has evolved
over different years, the metamorphic relations need to be updated to reflex those changes.
Table~\ref{tab:metamorphic-relations-changes} shows the modified metamorphic relations
with respect to the tax year 2020. These are some important changes: 
\begin{enumerate}
    \item Unlike previous years, married filing separately is qualified for $EITC$ as reflected in property \#3
    for the tax year 2021.
    \item The $AGI$ eligibility were different for different years as reflected in properties \#4, \#5, and \#6.
    \item In the version 2018 and 2019, the software does not allow users to check whether they itemize deductions.
    Therefore, we cannot specify the property \#14 for these two years.
    \item In the version 2018 and 2019, by not allowing for choosing itemized deductions, the software should automatically pick the option (standard vs. itemized) that maximizes the taxpayer benefits. Therefore, the modified property \#15 makes sure that software chooses standard deductions when the itemized deductions are lower. Similarly, property \#16 validates that the software uses itemized deductions when those deductions are higher. 
\end{enumerate}

We also observe that the U.S. Individual Income Tax Return is getting more
complicated each new year. For example, the software version 2018 takes 92 input fields,
whereas the version 2021 takes 187 input fields.

\begin{tcolorbox}[boxrule=1pt,left=1pt,right=1pt,top=1pt,bottom=1pt]
\textbf{Answer RQ1}: We found that metamorphic relations are suitable to specify the correctness requirements
in tax preparation software. Furthermore, these relations allow us to update the requirements as the tax policies
are evolving with minimal effort.
\end{tcolorbox}

{\footnotesize
\scriptsize
\begin{table*}[ht]
\caption{Experimental results on OpenTaxSolver using the $16$ metamorphic relations. Test-case generation time-outs at $15$ mins (results are averaged over 10 runs). }
\centering
\resizebox{\textwidth}{!}{%
\begin{tabu}{|l|llll|llll|llll|llll|}
\hline
\multirow{2}{*}{\textbf{Property ID}} & \multicolumn{4}{c|}{\textbf{OpenTaxSolver 2018}} & \multicolumn{4}{c|}{\textbf{OpenTaxSolver 2019}} & \multicolumn{4}{c|}{\textbf{OpenTaxSolver 2020}} & \multicolumn{4}{c|}{\textbf{OpenTaxSolver 2021}} \\
 &  \#\textit{test cases} & \#\textit{fail} & \#\textit{pass} & \textit{T$_{F}$(s)} &  \#\textit{test cases} & \#\textit{fail} & \#\textit{pass} & \textit{T$_{F}$(s)} & \#\textit{test cases} & \#\textit{fail} & \#\textit{pass} & \textit{T$_{F}$(s)} & \#\textit{test cases} & \#\textit{fail} & \#\textit{pass} & \textit{T$_{F}$(s)}  \\
\hline

Disability (1) & $36,558$ & $0$ & $36,558$ & $N/A$ & $35,970$ & $0$& $35,970$ & $N/A$ &  $36,255$ & $0$ & $36,255$ & $N/A$  & $32,456$ & $0$ & $32,456$ & $N/A$ \\
Disability (2) & $36,369$ & $0$ & $36,369$ & $N/A$ & $36,780$ & $0$& $36,780$ & $N/A$ & $35,790$ & $0$ & $35,790$ & $N/A$ & $32,355$ & $0$ & $32,355$ & $N/A$ \\
\hline
EITC (3) & $37,634$ & $0$ & $37,634$ & $N/A$ & $36,660$ & $0$& $36,660$ & $N/A$ & $19,936$ & $16,381$ & $3,555$ & $0.05$ & $32,343$ & $0$ & $32,343$ & $N/A$ \\
EITC (4) & $37,035$ & $0$ & $37,035$ & $N/A$ & $37,170$ & $0$& $37,170$ & $N/A$ & $18,258$ & $18,258$ & $0$ & $0.04$ & $16,556$ & $16,556$ & $0$ & $0.05$ \\
EITC (5) & $36,927$ & $0$ & $36,927$ & $N/A$ & $37,020$ & $0$& $37,020$ & $N/A$ & $36,450$ & $0$ & $36,450$ & $N/A$ & $32,883$ & $0$ & $32,883$ & $N/A$ \\
EITC (6) & $37,044$ & $0$ & $37,044$ & $N/A$ & $37,170$ & $0$& $37,170$ & $N/A$  & $36,360$ & $0$ & $36,360$ & $N/A$ & $32,962$ & $0$ & $32,962$ & $N/A$ \\
\hline
CTC (7) & $37,485$ & $0$ & $37,485$ & $N/A$ & $36,030$ & $0$& $36,030$ & $N/A$  & $36,120$ & $0$ & $36,120$ & $N/A$ & $32,388$ & $0$ & $32,388$ & $N/A$ \\
CTC (8) & $18,333$ & $0$ & $18,333$ & $N/A$ & $18,180$ & $0$ & $18,180$ & $N/A$ & $18,015$ & $0$ & $18,015$ & $N/A$ & $16,346$ & $0$ & $16,346$ & $N/A$ \\
\hline
ETC (9) & $37,548$ & $0$ & $37,548$ & $N/A$ & $37,020$ & $0$& $37,020$ & $N/A$  & $19,596$ & $15,744$ & $3,852$ & $0.05$ & $16,989$ & $15,886$ & $1,102$ & $0.05$ \\
ETC (10) & $36,081$ & $0$ & $36,081$ & $N/A$ & $36,720$ & $0$& $36,720$ & $N/A$  & $18,193$ & $18,166$ & $27$ & $0.06$ & $16,528$ & $16,494$ & $34$ & $0.05$ \\
ETC (11) & $17,334$ & $0$ & $17,334$ & $N/A$ & $18,450$ & $0$& $18,450$ & $N/A$  & $18,096$ & $33$ & $18,063$ & $33.40$ & $16,492$ & $33$ & $16,459$ & $29.02$ \\
ETC (12) & $18,486$ & $0$ & $18,486$ & $N/A$ & $18,390$ & $0$& $18,390$ & $N/A$ & $18,060$ & $0$ & $18,060$ & $N/A$ & $14,636$ & $0$ & $14,636$ & $N/A$ \\
\hline
ID (13) & $36,801$ & $0$ & $36,801$ & $N/A$ & $36,210$ & $0$& $36,210$ & $N/A$  & $36,160$ & $15$ & $36,145$ & $70.09$ & $27,348$ & $5,508$ & $21,840$ & $0.06$ \\
ID (14) & --- & --- & --- & --- & --- & --- & --- & --- & $36,405$ & $0$ & $36,405$ & $N/A$ & $31,916$ & $0$ & $31,916$ & $N/A$ \\
ID (15) & $36,926$ & $0$ & $36,926$ & $N/A$ & $36,630$ & $0$& $36,630$ & $N/A$ & $36,315$ & $0$ & $36,315$ & $N/A$ & $32,793$ & $0$ & $32,793$ & $N/A$ \\
ID (16) & $36,846$ & $0$ & $36,846$ & $N/A$ & $36,570$ & $0$& $36,570$ & $N/A$ & $36,235$ & $10$ & $36,225$ & $46.02$ & $32,363$ & $8$ & $32,355$ & $44.34$ \\
\hline
\end{tabu}
}
\label{table:experiments}
\end{table*}
}

\subsection{Test-Case Generations (RQ2)}
We implement the Algorithm~\ref{alg:random-testing} in \toolname to generate base and follow-up test cases and find the evidence that
the software fails to satisfy the metamorphic requirements. In doing so, we find inputs that satisfy the premises, but it negates the conclusion of
metamorphic relation. We run each experiment $10$ times, each for $15$ minutes (the results are averaged over these repeated runs).

Table~\ref{table:experiments} shows the results of random test-case generations.
We organize the table into four parts for the tax year 2018, 2019, 2020, and 2021, respectively. As we
discussed previously, we cannot specify the metamorphic relation for ID (14), hence we skip this property for testing
of software versions 2018 and 2019. 
We report the number of generated test cases, the number of passed ones, the number of
failed cases, and the time to observing the first failed case (in seconds).
Overall, we found that our search strategy is efficient that generates over $36,000$ test cases for $2$-property relations and $17,0000$ for $4$-property relations in average. We also found that the testing technique can often find the first evident of failures within a few seconds and up to $70$ seconds in the worst-case.

\vspace{0.5em}
\noindent\textbf{Different Variant of Software.} As the tax law evolves in the response to
ever-changing politico-economic realities, the tax preparation software has to
be adapted to reflect those changes. Our experiences with metamorphic testing
show that the tax preparation software might not be properly updated to account for
those changes. In particular, we did not find any evidence of failures for all $16$
properties for 2018 and 2019 versions, whereas there is evidence of failures
for 2020 and 2021 versions. One specific example is property \#4
where the software satisfies the property for 2018 and 2019, but it fails for 2020.
Upon further investigation, it seems
that the software has not been updated to reflect the new eligibility based on the annual
growth income (see Id=4 in Table~\ref{tab:metamorphic-relations-changes}).
Another interesting finding is for property \#3, comparing the tax years
2020 and 2021. For the tax year 2020, the status of married filing separately (MFS)
was not eligible for earned income tax credits (EITC); however, this status has become
eligible for EITC in tax year of 2021. While, the 2020 version failed
to satisfy property \#3, the 2021 version (without any updates) satisfies this
property, solely due to the updated policy. 

\vspace{0.5em}
\noindent\textbf{Failed Metamorphic Relations.} We found test cases that violate
the correctness requirements for the 2020 and 2021 versions of software.
One area that software shows weakness is the filing status of married filing
separately (MFS) where two properties (properties \#3 and \#9) out of three correctness
specifications for this status (properties \#3, \#7, and \#9) are failed. While,
the majority of filing are single or married filing jointly, the MFS status
is still an important class of filing for taxpayers in the US when
3.07 million people used this status to file their taxes in 2016 (out of
150.3 million federal returns filed). Another area of vulnerability is
the annual growth income (AGI) that influences the correctness of
software in many specifications (9 properties out of 16 have some
conditions on the AGI). Since the eligibility based on AGI has consistently
been modified by IRS in different years (see Table~\ref{tab:metamorphic-relations-changes}),
it is critical to take into account a design that requires minimal efforts
to update conditions for different computations (e.g., EITC, CTC, and ETC). 

\begin{tcolorbox}[boxrule=1pt,left=1pt,right=1pt,top=1pt,bottom=1pt]
\textbf{Answer RQ2}: We found that our proposed random search algorithm is effective and efficient in exploring the space of tax preparation software to uncover failures. Our experiences revealed multiple areas where updated software is no longer satisfying the correctness requirements. In addition, our experiments showed multiple weakness areas relate to married filing separately status and the maximum annual growth income.
\end{tcolorbox}

\begin{figure*}[!htb]
    \centering
    \begin{minipage}{0.32\textwidth}
        \centering
        \includegraphics[width=1.0\textwidth]{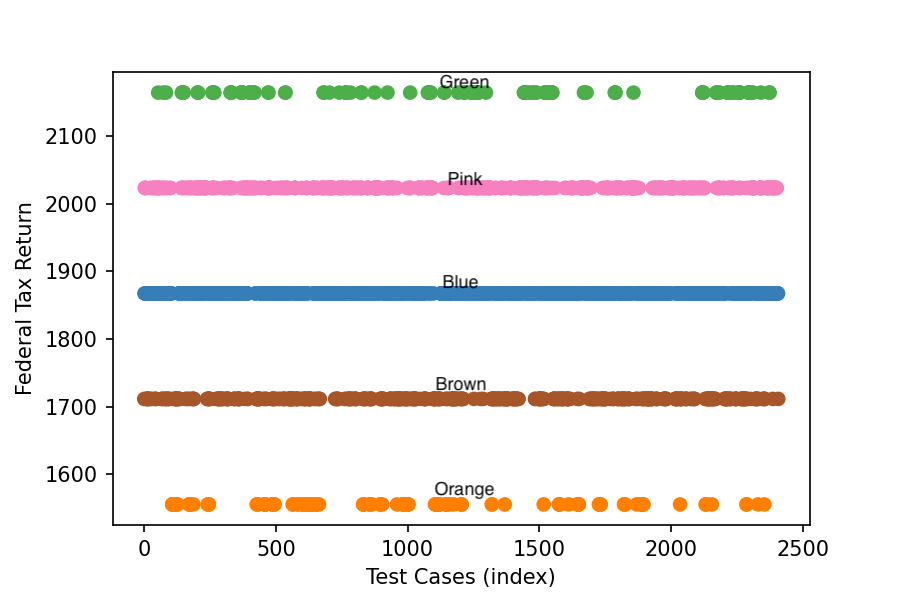}
    \end{minipage}%
    \begin{minipage}{0.66\textwidth}
    	\centering
        \includegraphics[width=1.0\textwidth]{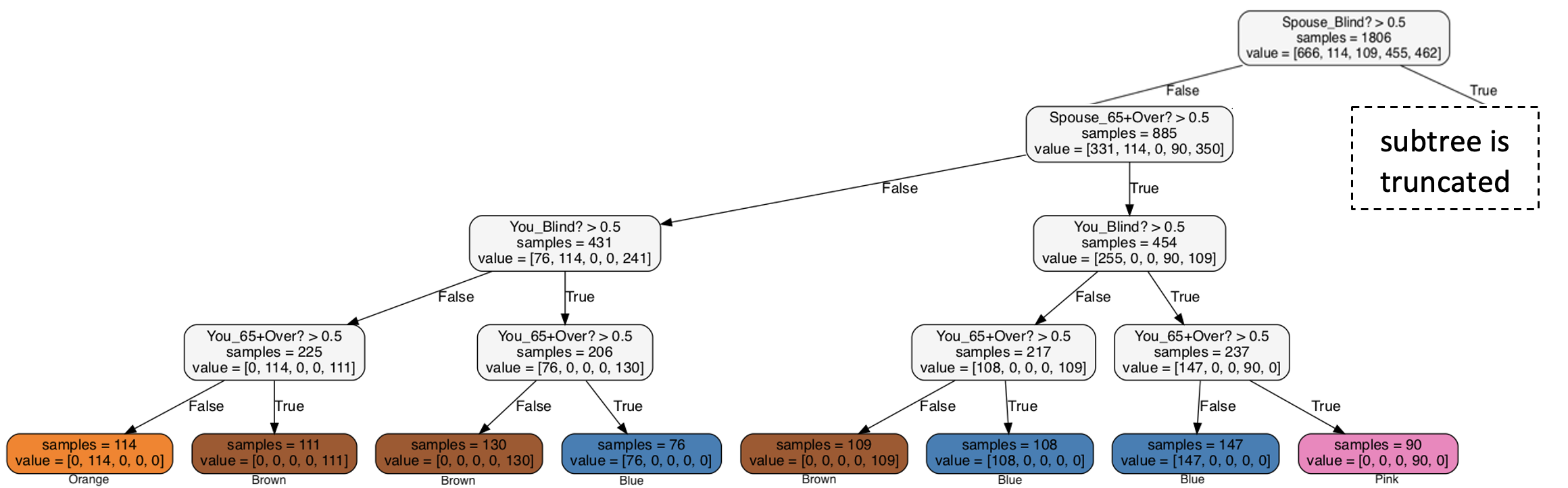}
    \end{minipage}
    \caption{
    \textbf{Left.} Tax returns are clustered into $5$ groups, varying values of variables
    in properties \#1,\#2;
    \textbf{Right.} DT explains the number of \textit{yes} answers to four senior/disability
    (age above 65, blindness, spouse age over 65, and spouse blindness) explains different clusters of Federal Tax Return.}
    \label{fig:case-study-1}
\end{figure*}

\begin{table}[h!]
      \caption{Debugging in the input and internal space.}
      \label{tab:debugging}
     \begin{center}
     \begin{tabular}[t]{ p{0.5cm} | c | c }
     \toprule
      Id & Debugging Input Space & Debugging Internal Space 
      \\    \midrule
     EITC (3) &
     {\includegraphics[width=0.19\textwidth]{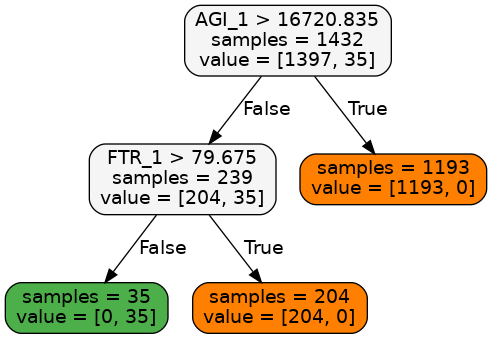}}
      & 
     {\includegraphics[width=0.19\textwidth]{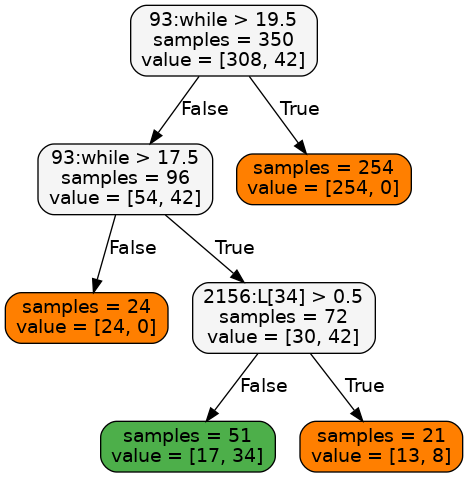}}
      \\    \midrule
     ETC (9) &
     {\includegraphics[width=0.19\textwidth]{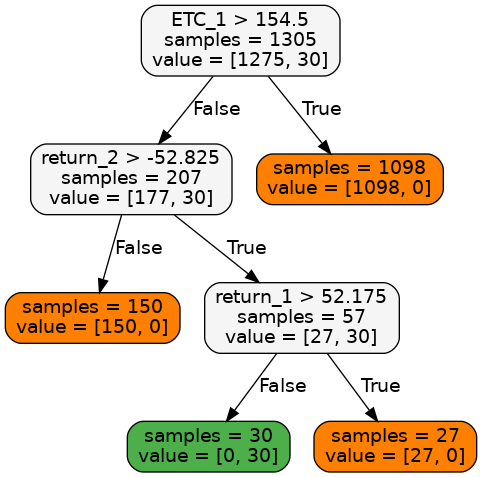}}
      & 
     {\includegraphics[width=0.19\textwidth]{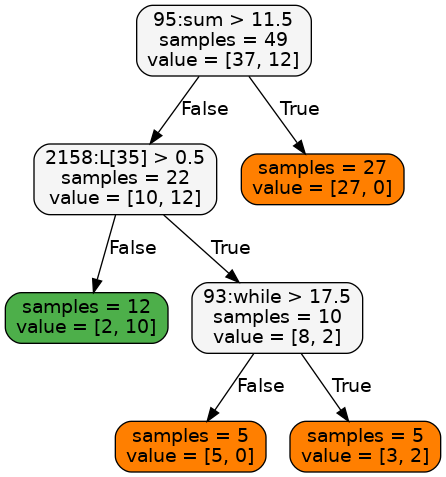}}
      \\    \midrule
     ETC (11) &
     {\includegraphics[width=0.19\textwidth]{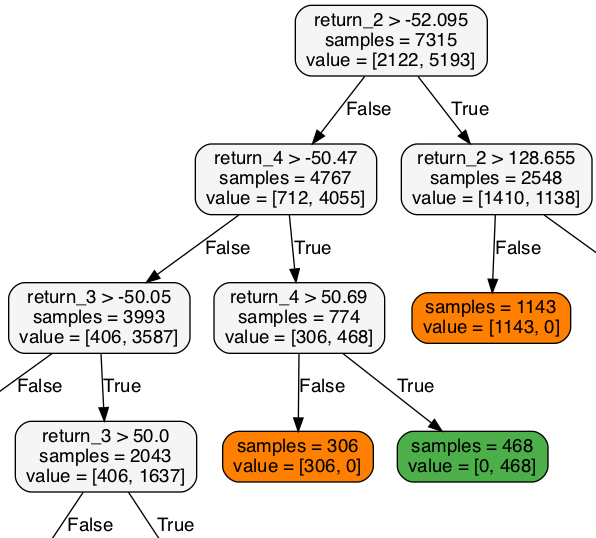}}
      & 
     {\includegraphics[width=0.19\textwidth]{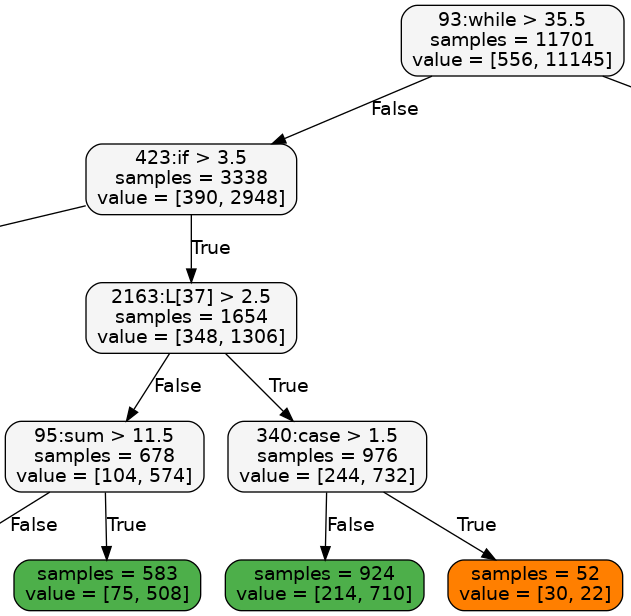}}
      \\    \midrule
     ID (13) &
     {\includegraphics[width=0.19\textwidth]{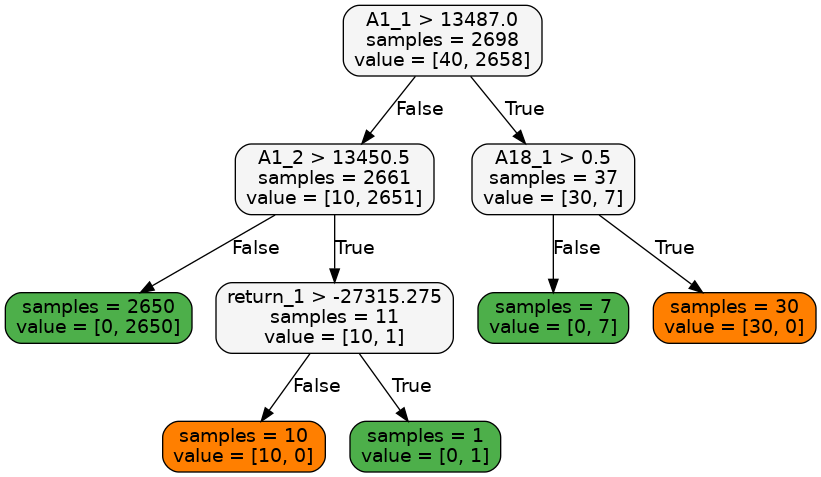}}
      & 
     {\includegraphics[width=0.19\textwidth]{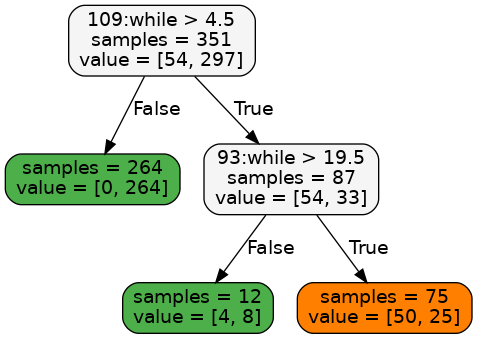}}
      \\    \midrule
     ID (16) &
     {\includegraphics[width=0.19\textwidth]{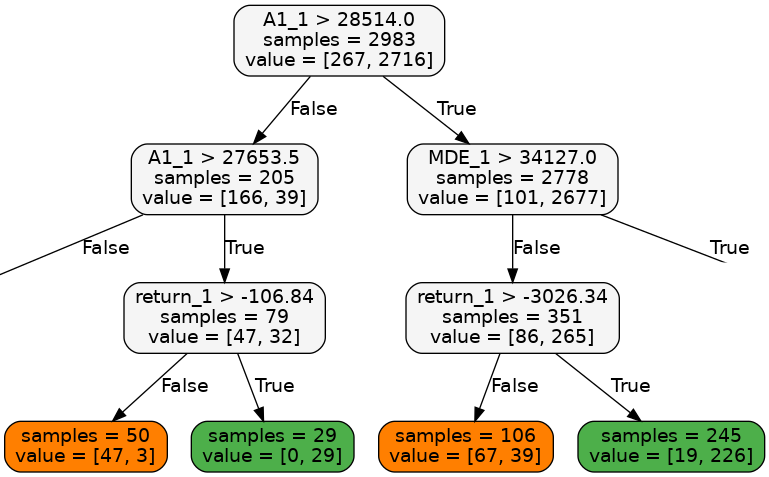}}
      & 
     {\includegraphics[width=0.19\textwidth]{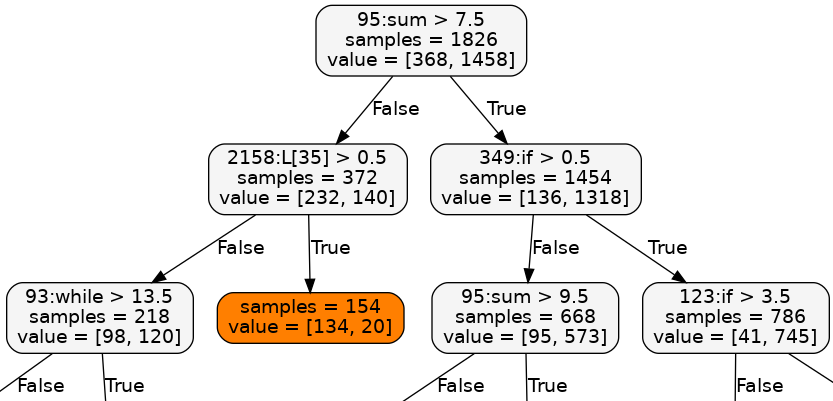}}
     \\ \bottomrule
      \end{tabular}
      \end{center}
      \vspace{-1.0em}
    \end{table}

\subsection{Fault Localization (RQ3)}
In this section, we investigate whether decision tree artifacts
are expressive and useful to aid tax preparation software developers
and users in explaining and debugging different correctness properties.
We first use a simple example of disability (properties \#1 and \#2)
to showcase the usefulness of decision tree artifacts. Then, we focus
on the properties that have failed to satisfy the metamorphic correctness
requirements to explain and localize root causes. 

Figure~\ref{fig:case-study-1} (left) shows five clusters of 
federal tax returns (FTR) obtained by varying four disability/senior
conditions (age above 65, blindness, spouse age over 65, and spouse blindness).
Since there are $16$ possibilities for these four Boolean conditions, our goal
is to infer a decision tree model that explains what conditions are similar
within each cluster and what conditions distinguish different clusters
from each other. Figure~\ref{fig:case-study-1} (right)
shows the corresponding DT model for explanations. Each path from root nodes to the
leaf defines a condition on inputs such that if any input satisfies the condition,
it assigns to the corresponding cluster label of leaf node.
For example, the lowest orange cluster (FTR below \$1,600)
is explained by a condition where their age and their spouse's age are less than $65$,
and none of them are blind.  The second cluster from below (Brown color with the FTR
a bit above \$1,700) is explained by four paths where the taxpayer
has one disability/senior condition out of four possibilities. The third cluster
from below (blue) is described by six paths where the taxpayer has two
disability/senior conditions. In a similar fashion, the next two clusters
have three and four disability/senior conditions out of four possibilities.
Therefore, the decision tree shows that the number of ``yes'' answers to
four disability/senior conditions (that can be 0, 1, 2, 3, and 4) explains
the five clusters found in the Federal Tax Return.

\vspace{0.5em}
\noindent\textbf{Explanations in the input space.}
\toolname uses a similar technique to explain circumstances on the input variables
(different items in the tax return) under which the software
fails to satisfy the metamorphic properties. We label
failed test cases with orange colors and passed ones with green colors. 
Note that if all test cases are failed, we just report the premises
of the corresponding metamorphic property, and it means the software completely
miss those conditions. Otherwise, we provide debugging supports via decision tree artifacts.
Table~\ref{tab:debugging} (second column)
shows explanations in the input space of software. We note that
features ending with \_1 are bases and \_2 are follow-ups (for 4-properties,
features ending with \_1 and \_2 are bases and \_3 and \_4 are follow-ups).

For EITC (3), the decision tree shows that the software fails in all cases excepts
when the annual growth income (AGI) is less than \$16,720 and the federal
tax return (FTR), shown with return, is less than \$79. In other words, the software fails to
enforce property \#3 in Table~\ref{tab:metamorphic-relations} in all cases,
except for cases with a federal tax return of less than \$80. 
For ETC (9), the decision tree shows that the software also fails to satisfy
the metamorphic requirements in all cases, except when the tax return becomes zero.
We conjecture that the correct cases are due to numeric
approximations as the computations are expected to use a small non-zero value,
but are approximated due to the finite precision. Therefore, the software is correct
for those cases solely because of numerical errors. Similarly, for ETC (11) and ID (16),
the software satisfies the properties in the majority of cases but fails in a few corner cases.
Inspecting these failures, we found that they also include an interval
of FTR with zero return. So, the software behaves erroneously due to numerical approximations,
but is correct otherwise. For ID (13), the decision tree
model shows circumstances over the medical expenses (the first line of schedule A shown with A1
and MDE variable) and the return. The software fails when the medical
expenses are over \$13,487 and the tax payer opts to itemize
deductions (A18 is set to \texttt{true}). In other words, the software behaves correctly if the
user does not itemize, but becomes faulty when the user opts for the itemized deduction.
We conjecture that this is
due to the fact that the 2020 version was updated to allow users to opt for standard
versus itemized deductions, which were not possible in past versions.
But, the updates might not be properly implemented for all corner cases and let users take deductions
when they opt for itemized deduction, but do not meet the eligibility conditions.

\vspace{0.5em}
\noindent\textbf{Explanations in the internal space.}
We further extend \toolname to aid tax preparation
software developers to debug their code. In particular,
we collect internal traces of each execution such as how many times
a \texttt{loop} is executed, what branch in an \texttt{if} condition is taken, and what
case in a \texttt{switch} statement is executed. We use the same label
as the debugging in the input space where each traces is labeled
as passed (green) and failed (orange), and we want to localize
program internal properties that distinguish failed traces from passed ones.
The decision trees learned over program internal features are shown in
the third column of Table~\ref{tab:debugging}. 

As an example, let us consider EITC (3). The failed and passed test
cases are distinguished based on the number of iterations of \texttt{while} loop
in line 93 and whether the tax computation ends up refunding a return
to the user in line 2156 with the variable L[34]. For another example,
the decision tree  model for ETC (11) pinpoints to an \texttt{if} condition
and a \texttt{switch} statement in line 423 and 340, respectively.

\begin{tcolorbox}[boxrule=1pt,left=1pt,right=1pt,top=1pt,bottom=1pt]
\textbf{Answer RQ3:} 
We found that decision trees are useful artifacts to explain circumstances
under which the tax preparation software fails to satisfy the metamorphic
requirements. Our experiences reveal that (i) the software might completely
miss an eligibility condition, (ii) finite precision in the computation
might lead to unexpected errors, and (iii) the updated version allows
users to opt for an option might not consider all cases, especially the corner ones.
\end{tcolorbox}

\section{Related Work}
\label{sec:related}
\noindent \textbf{Tax Preparation Software.}
Merigoux et al.~\cite{10.1145/3446804.3446850} developed a compiler for
the French tax code. They pointed out multiple limitations of 
$M$ programming language, developed by French Public Finances Directorate
to write tax rules (available at~\cite{French-M-Rules}).
Then, they developed a domain-specific language that allows for specifying
complicated rules and lifting them to modern languages like Python.
They also used a dynamic random search to validate the software, as opposed
to formal verification, due to the large space of inputs, floating-point computations, and various optimizations. In comparison, the US has much less automation in tax preparation with no available specifications and regulated test cases. Yu, McCluskey, and Mukherjee~\cite{yu2020tax} proposed knowledge-based graphs to personalize the tax preparation within the TurboTax software. In addition, SARA~\cite{DBLP:conf/kdd/HolzenbergerBD20} translated the statutes into Prolog programs and the cases into Prolog facts, such that each case can be answered by a single query. Alternatively, they proposed to adapt high-dimensional natural language processing techniques such as Legal BERT to overcome some of intrinsic limitations of logic-based programming. 

\noindent \textbf{Fairness.}
Algorithmic fairness is an important area to promote inclusion, diversity, and accessibility of software solutions~\cite{10.1145/3106237.3106277,10.1109/ICSE43902.2021.00129,10.1145/3510003.3510202}. Within the tax domain, Black et al.~\cite{10.1145/3531146.3533204} studied fairness of algorithmic tax audit selections by the United States Internal Revenue Service (IRS) from 2010-14 using the concept of vertical equity. They found that flexible machine learning with higher accuracy, as opposed to simpler ones, may undermine vertical equity by shifting tax audit burdens from high-income to middle-income taxpayers. They also pointed out multiple weaknesses in applying existing fairness solutions to the tax audit problems. Finally, they made multiple suggestions to improve the vertical equity through fair audit selection. Other articles also made similar observations that the IRS might audit low-income taxpayers at the same rate as the richest taxpayers~\cite{Bias-tax-notes,IRS-ProPublica-article,IRS-theatlantic-article}. For example, a ProPublica article reported that the top 1\% of taxpayers by income were audited at a rate of 1.56\% whereas earned income tax credit (EITC) recipients, who typically have an annual income under \$20,000, were audited at 1.41\% in 2018. Multiple works also pointed out the issue of ``color-blind'' tax code in the US~\cite{Moran98,Bearer-Friend18,Bearer-Friend21}. Moran~\cite{Moran98}
used the horizontal equity as fairness notion that requires similarly situated taxpayers should
have taxed similarly. The study found that blacks and whites who are similar by
income are not taxed similarly because they fundamentally have different lifestyles. Studying fairness and ethical aspects of tax law as implemented in a tax preparation software system is an interesting future direction. 

Beyond the US tax system, a Dutch digital welfare fraud detection system, known as SyRI, is ruled to be unlawful since it does not comply with the
right to privacy under the European Convention of Human Rights and the European Union
General Data Protection Regulation (GDPR).
Furthermore, the court found that the SyRI legislation is insufficiently transparent
and verifiable and brings risks of discrimination~\cite{doi:10.1177/13882627211031257}.

\section{Broader Impacts}
\label{sec:broader-impacts}

Although most of our technical discussion focused on the US-based tax preparation software,
the key takeaway is on the importance of metamorphic relations for the sociolegal software. 
We discuss two examples highlighting this connection.

\noindent {\bf Legal-Critical Software.} 
Let us first consider the work by Matthews el al.~\cite{10.1145/3306618.3314279,10.1145/3375627.3375807} on
forensic DNA software that aims to understand the role of black-box forensic
software in moral decision-making in criminal justice. They conduct independent
testing of Forensic Statistical Tool (FST), a forensic DNA system developed in
2010 by New York City’s Office of Chief Medical Examiner (OCME). Using a collection of over 400 mixed DNA samples, they found that an undisclosed data-dropping method
in software impacts about 25\% of the samples and leads to shifted results
toward false inclusion of individuals who were not present in an evidence sample.
The techniques developed in our paper may allow one to uncover such vulnerabilities without having an access to DNA samples purely on the basis of an appropriate metamorphic query uncovering differential data-dropping.

\noindent  {\bf Socio-Critical Software.} Escher and Banovic~\cite{10.1145/3392874} studied errors in poverty management systems that are critical tools to support  vulnerable populations. Online benefits screening tools advise households about their eligibility before proceeding with full applications. Hence, an error in such a system can deprive qualified families from receiving the benefits they are entitled to. Escher and Banovic developed a framework to test an instance of such screening tools as implemented in the Pennsylvania ``Do I Qualify?'' website~\cite{COMPASS-HHS}. They took the benefit eligibility handbook~\cite{Penn-Human} and implement their own screening software. Then, they use census data to generate test households and compare the results of the online tool against their own implementations. Taking their own implementations as the ground truth, they mark any discrepancies as errors. Overall, they found major errors in the implementation of the tool that can be used as necessary correctives to fix them to improve accessibility for the most vulnerable populations. Metamorphic testing again can complement this approach by eliminating the need for a comparative implementation. 

\section{Conclusion}
\label{sec:conclusion}
Tax preparation software are increasingly indispensable in navigating the complex tax structure. Since the legal responsibilities of bugs in such software rest on the end-users, 
most of whom cannot afford expensive accountants or commercial software, 
developing debugging aids for such systems is our social responsibility. 
We presented a data-driven debugging framework for metamorphic specifications expressed in a first-order logic.
We hope that our experience in capturing the tax-preparation software requirements can convince the reader that the metamorphic specifications are both intuitive and unambiguous in expressing the comparative nature of legal requirements. 
The experimental results demonstrated that search-based software engineering approaches are useful in uncovering and explaining bugs in a popular open-source tax preparation software.


\vspace{1.0 em}
\noindent \textbf{Acknowledgement.}
We would like to thank Nina Olson (J.D), the Executive Director of the Center for Taxpayer Rights, to discuss this paper with us and provide useful suggestions to improve the paper. The authors would like to also thank anonymous ICSE-SEIS reviewers for their time and valuable feedback. Finally, Tizpaz-Niari acknowledges the start-up supports from the UTEP College of Engineering in carrying out this research.
\vspace{1.0em}

\bibliographystyle{IEEEtran}
\balance
\bibliography{main}

\end{document}